\documentclass[prb,aps,amssymb,twocolumn,showpacs,superscriptaddress]{revtex4}
\usepackage{graphicx}
\usepackage{amsmath}
\usepackage{amssymb}
\usepackage[latin1]{inputenc}

\begin{document}

\title{Geometry-related magnetic interference patterns in long SNS Josephson junctions}

\author{F.~Chiodi}
\author{M.~Ferrier}
\author{S.~Guéron}
\affiliation{Laboratoire de Physique des Solides, UMR 8502, CNRS, Université Paris-Sud, 91405 ORSAY Cedex, France}
\author{J. C.~Cuevas}
\affiliation{Departamento de F\'isica Téorica de la Materia Condensada, Universidad Aut\'onoma de Madrid, 28049 Madrid, Spain}
\author{G.~Montambaux}
\affiliation{Laboratoire de Physique des Solides, UMR 8502, CNRS, Université Paris-Sud, 91405 ORSAY Cedex, France}
\author{F.~Fortuna}
\affiliation{Centre de Spectrométrie Nucléaire et de Spectrométrie de Masse, UMR 8609, CNRS, Université Paris-Sud, 91405 ORSAY Cedex, France}
\author{A.~Kasumov}
\author{H.~Bouchiat}
\affiliation{Laboratoire de Physique des Solides, UMR 8502, CNRS, Université Paris-Sud, 91405 ORSAY Cedex, France}

\date{\today}
\begin{abstract}
We have measured the critical current dependence on the magnetic flux of two long SNS junctions differing by the normal wire geometry. The samples are made by a Au wire connected to W contacts, via Focused Ion Beam assisted deposition. We could tune the magnetic pattern from the monotonic gaussian-like decay of a quasi 1D normal wire to the Fraunhofer-like pattern of a square normal wire. We explain the monotonic limit with a semiclassical 1D model, and we fit both field dependences with numerical simulations of the 2D Usadel equation. Furthermore, we observe both integer and fractional Shapiro steps. The magnetic flux dependence of the integer steps reproduces as expected that of the critical current $I_c$, while fractional steps decay slower with the flux than $I_c$. 
\end{abstract}
 \maketitle

\begin{figure}
\includegraphics[width=\columnwidth]{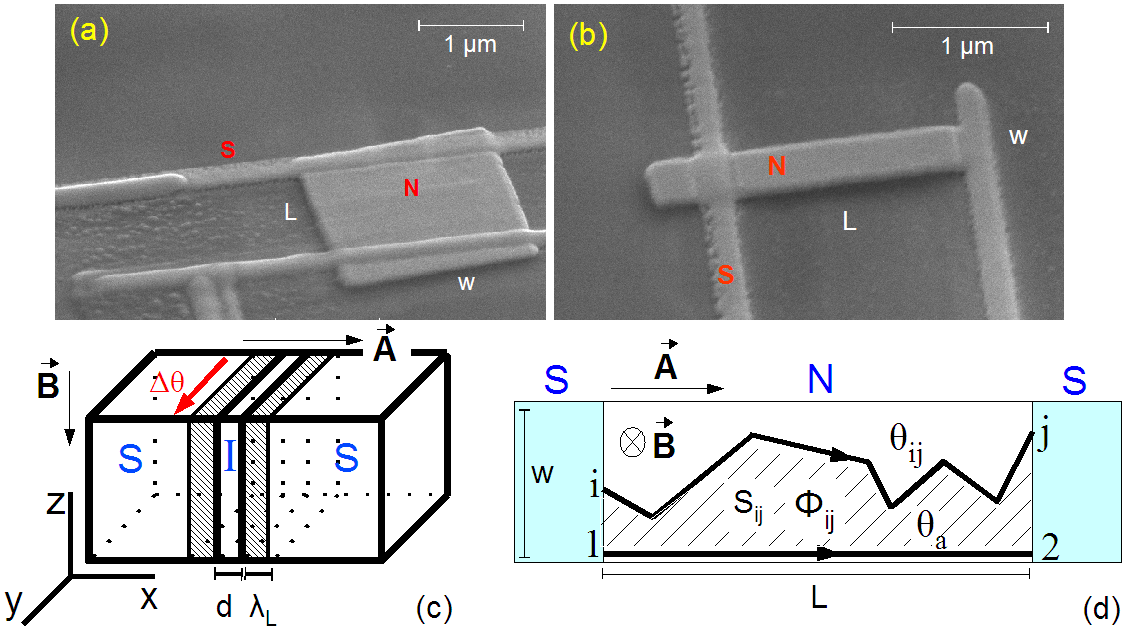}
\caption{(color online) Scanning Electron Microscope images of samples WAu-Sq (a) and WAu-N (b). One can see as a clear halo the W contamination around the superconducting contacts. (c) Scheme of a SIS Josephson junction. (d) Scheme of the 1D model developped. }
\label{semNYgilles}
\end{figure}

\emph{Introduction --}
A non-dissipative supercurrent can be transmitted between two superconductors (S) through a nanometer-thin insulating layer (I), when a phase difference is imposed. Superconducting correlations can also penetrate into a micrometer-long, non-superconducting coherent metal (N) over lengths greater than $\xi_S$, the superconducting coherence length in N. A supercurrent can then flow through long ($L>\xi_s$) SNS junctions, provided that the phase coherence is preserved in the normal wire. The supercurrent amplitude only depends on the normal metal length $L$ and diffusion coefficient $D$ (which set the characteristic Thouless energy $E_{Th}=\hbar D/L^2$) and on the normal metal resistance. \cite{heikkila} The supercurrent thus reflects the transport mechanisms in the normal wire, and is affected by the interference and diffraction phenomena present in the normal metal as a result of the phase coherence.\\   
We have studied the behavior of the supercurrent in long SNS junctions in a perpendicular magnetic field. In a previous work, \cite{Lionel} we observed that the maximum supercurrent, the critical current $I_c$, monotonously decreased with the magnetic field following a quasi-gaussian dependence. This behavior is different from the interference Fraunhofer pattern usually observed, for example, in SIS junctions, short SNS junctions ($L \le \xi_s$), magnetic SFS Josephson junctions, etc. 
This difference is induced by the different aspect ratios of short weak links and long SNS junctions. Indeed, in SIS junctions the thickness of the junction is limited to a few Angstroms, to permit the tunneling of Cooper pairs; similarly, in SFS junctions the magnetic layer thickness has to be shorter than a few tens of nanometers, so that Cooper pairs are not broken by the internal exchange field. In contrast to the wide and short SIS and SFS junctions, SNS junctions offer the interesting possibility to explore a broad range of aspect ratios, since the length of the normal metal is only limited by the phase coherence length, which can be as long as a few microns at low temperature. In this paper we explore different geometries of long SNS junctions. Both monotonic and non-monotonic $I_c(B)$ dependences have been observed before, but we show for the first time that we can tune the $I_c (B)$ curve from an interference pattern to a quasi-gaussian monotonic dependence by varying the normal metal's aspect ratio.\\
Monotonic $I_c (B)$ dependences have been observed in ballistic long SNS junctions, formed by a normal bidimensional InAs electron gas connected to superconducting Nb contacts larger than the London screening length. \cite{Strunk} Their magnetic field dependence resulted from the 
screening currents in the Nb. On the contrary, the diffusive junctions investigated in the present work are contacted by thin disordered superconducting  wires in which the  magnetic field screening is negligible. We show that the geometry dependent magnetic field decay can be explained taking into account only the interferences between Andreev pairs' trajectories in the normal metal.\\  
As a reminder, we first consider the case of a wide short SIS junction. A magnetic field in the SIS junction plane penetrates in the insulating layer of thickness $d$ and in the superconductors nearby over a length $\lambda_L$, the London penetration length. In a magnetic field $\vec{B} = - B \hat{z}$ of vector potential $\vec{A} = B y \hat{x}$, the phase shift of the Cooper pairs tunneling at different points of the junction width is (Fig. \ref{semNYgilles}(c)):
\begin{equation}
 \theta (y)= 2 \pi \frac{2 e}{h}\int_{-\lambda_{L}}^{d+\lambda_{L}}{A_x dx} = 2 \pi \frac{\Phi(y)}{\Phi_0}
\end{equation}
where $\Phi (y)$ is the flux through the surface $S_y = (d+2\lambda_L) \, y$, and $\Phi_0 = h/(2e)$ is the quantum flux.
The current is obtained by integrating over the junction surface the supercurrent density $j = j_c\, sin(\delta+\theta)$, taking into account both the superconducting phase difference between the contacts $\delta$ and the phase due to the vector potential. The critical current dependence on the magnetic flux $\Phi=B (d+2\lambda_L) w$, follows the well-known Fraunhofer pattern, a diffraction pattern created by the interference between the ballistic trajectories over the junction width:
\begin{equation}
	I_c = I_c(0) \, \frac{\Phi_0}{\pi \Phi} \, \bigg\vert \sin \bigg( \frac{\pi \Phi}{\Phi_0}  \bigg)  \bigg \vert
\end{equation}
If we now consider SNS junctions, we expect strong differences between junctions containing a short and wide or a long and narrow normal wire: in a short wide wire the phase difference between the trajectories comes from the phase distribution along the junction width, while in a narrow long wire, the phase of each trajectory is accumulated along the junction length.\\

\emph{The samples --} We have fabricated long SNS junctions where a Au normal wire links two superconducting W contacts. First, a $50\, nm$ thick Au wire is drawn by e-beam lithography and deposited onto a Si0$_2$ substrate. We use $99.9999\%$ pure gold, with a content in magnetic impurities (Fe) smaller than 0.1 ppm. This insures a long phase coherence length $L_{\Phi} \sim 10 \,\mu m$ below 50 mK, measured in a separated weak localisation experiment. We then deposit the superconducting contacts in 
a Focused Ion Beam (FIB). After slightly etching the Au wire with the FIB to remove possible impurities on its surface, we inject a metallo-organic vapor of tungsten hexacarbonyl over the sample. This vapor is decomposed by the focused Ga$^+$ ion beam, and a disordered W alloy is deposited on the substrate. The wires produced are composed of tungsten, carbon and gallium in varying proportions (in our case, the atomic concentrations are roughly 30$\%$ W, 50$\%$ C and 20$\%$ Ga). \cite{Sandrine} The superconducting critical temperature of the wires produced is $T_c \sim 4\,$K, an order of magnitude higher than the bulk $T_c$ of W. \cite{W} This could be due to the inclusion of Ga, which is itself a superconductor with $T_c=1\,$K. The critical magnetic field of the wires is also strikingly high: at $1\,$K, $H_c=7\,$T. \cite{AlikFIB} The W wires are 200 nm wide and 100 nm thick. The dependence of the superconducting properties of these wires on the deposition conditions have been investigated in detail in W. Li {\it{et al.}}. \cite{li} The superconducting gap as well as the Abrikosov flux lattice have been studied by STM experiments. \cite{vieira} The investigation of proximity induced superconductivity in metallic  nanowires contacted by FIB has also been recently performed. \cite{FIBusers}
The advantages of this technique are the deposition of the material of practically any shape and size, without any mask, and the good quality of the interface created. The main disadvantage is the Ga contamination of about 250 nm around the deposited wires. The long SNS junctions created by FIB-assisted deposition are comparable to junctions created with more standard fabrication methods. We thus recovered the general results for the voltage vs. current curves and the temperature dependence of the critical current. \cite{DubosIcT}  \\
To investigate the influence of the geometry on the $I_c(\Phi)$ dependence, we have designed two samples with different aspect ratios: sample WAu-Sq is 1.2 $\mu$m long and 1.75 $\mu$m wide, with an aspect ratio $L/w=0.7$, while sample WAu-N is 1.53$\mu$m long and 0.34 $\mu$m wide, with an aspect ratio $L/w=4.5$ (Fig. \ref{semNYgilles}(a),(b)).

\begin{figure}
\includegraphics[width=\columnwidth]{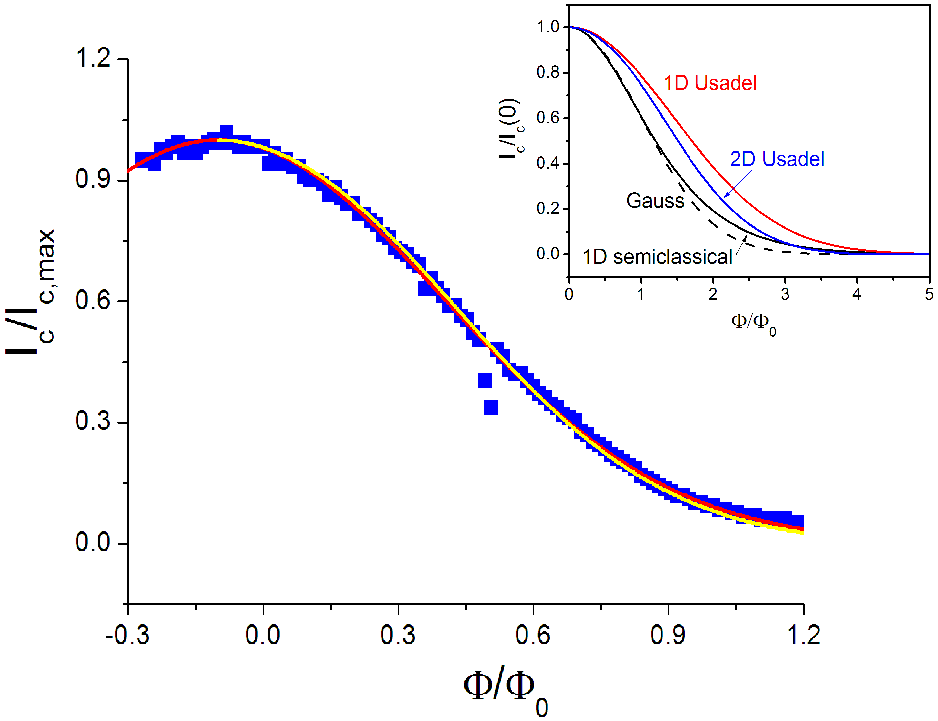}
\caption{(color online) Sample WAu-N normalised critical current vs. normalised flux at T=60 mK. Red line: Gaussian function with $\sigma$=0.5. Yellow line: numerical simulation of the 2D Usadel equation, where the flux has been rescaled by a factor 2.5.  Inset: comparison between a Gaussian dependence (dotted line), the semiclassical model prediction for a 1D normal wire (black line), the analytical result of Usadel equation in the limit $L>>w$ (red line) and the numerical simulation of the 2D Usadel equation for the aspect ratio of sample WAu-N (blue line). $I_{c,max}$=2.4 $\mu$A.}
\label{IcHN}
\end{figure}

\begin{figure}
\includegraphics[width=\columnwidth]{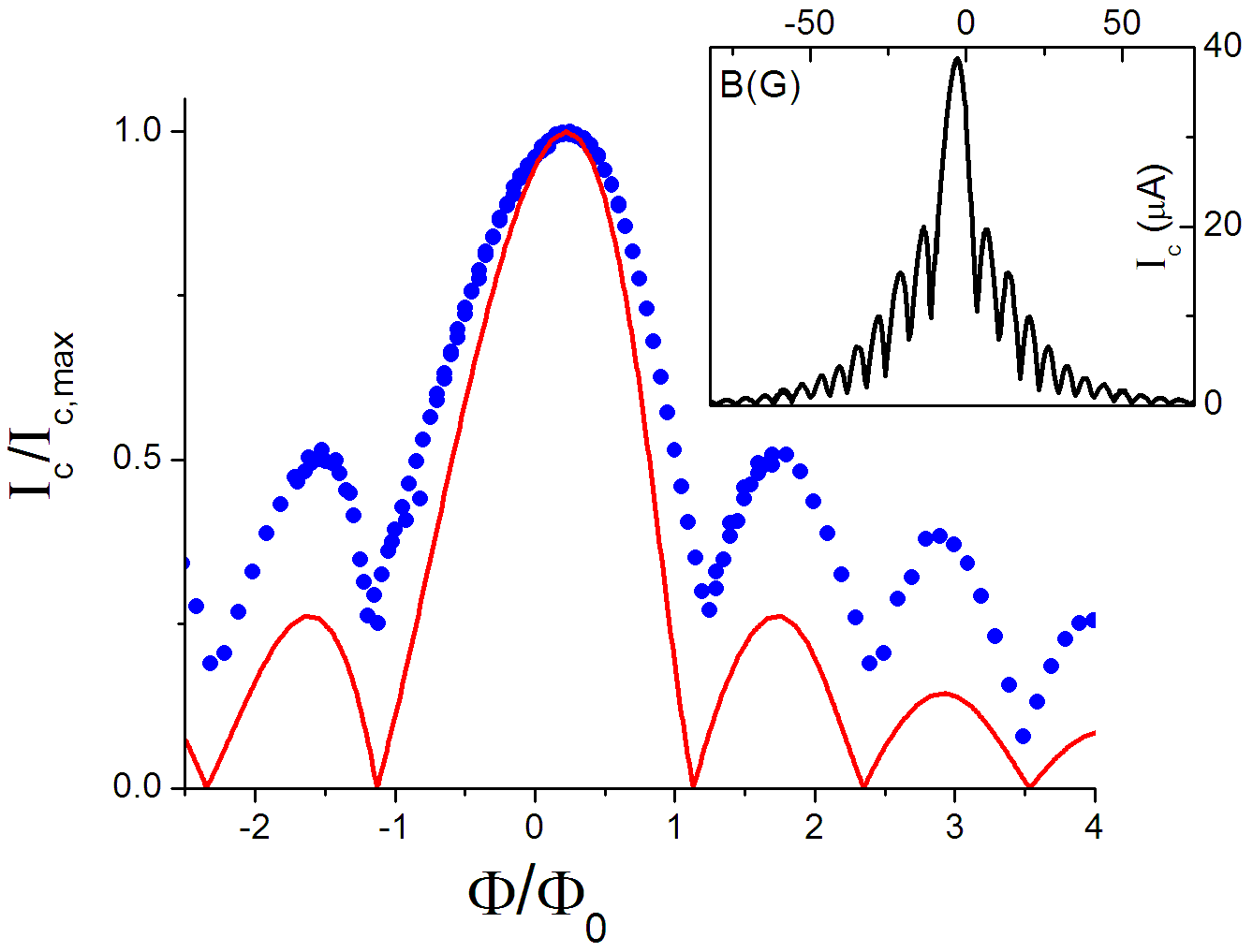}
\caption{(color online) Sample WAu-Sq normalised critical current vs. normalised flux (blue dots). A trapped flux of 4.3 G has been subtracted. Red line: numerical simulation of the 2D Usadel equation for a junction with aspect ratio $L/w$=0.7 and W wires inductance $\mathcal{L}$=11.5 pH. Inset: raw data for $I_c(B)$, with $I_{c,max}=$38.8 $\mu$A. T=60 mK.}
\label{IcHY}
\end{figure}

The $I_c(\Phi)$ normalised curves for samples WAu-N and WAu-Sq are shown in Fig. \ref{IcHN} and \ref{IcHY} respectively. They illustrate the important role of the aspect ratio: sample WAu-N displays a quasi-gaussian decay of the critical current, while sample WAu-Sq displays oscillations which recall a Fraunhofer pattern.\\

\emph{Semiclassical model --} To explain the behavior of sample WAu-N, a semi-classical model was developed. \cite{Montambaux} The aim is to model a long diffusive SNS junction, with $w<<L<<L_\phi, L_T$ (1D geometry), in a perpendicular magnetic field $\vec{B} = -B \hat{z}$ of vector potential $\vec A = B \,y\, \hat{x}$. We suppose a sinusoidal current-phase relation for simplicity: $j = j_c \, \sin(\delta + \theta_{ij})$. $\theta_{ij}$ is the phase acquired along the trajectory starting at point $i$ and finishing at point $j$. $\delta$ is the phase difference between the two superconducting contacts, and is independent of the position of $i$ and $j$, since in this long 1D geometry one can ignore the dephasing along the junction width in comparison to the one along the junction length. The total current is the average over all possible trajectories $\mathcal{C}_{ij}$:
\begin{equation}
	I \propto \mbox{Im} \bigg [<e^{i(\delta + \theta_{ij})}>_{\mathcal{C}_{ij}} \bigg]
\end{equation}
Considering a reference trajectory 1-2 with a phase difference $\theta_a$, we have $\theta_{ij} = \theta_a +\Delta \theta_{ij}$ (see Fig. \ref{semNYgilles}(d)). The critical current is the maximum of the current $I$:
\begin{equation}
	I_c \propto \mbox{Im} \bigg [ e^{i(\delta + \theta_{a})}<e^{i \Delta \theta_{ij}}>_{\mathcal{C}_{ij}} \bigg]_{max} \propto  \bigg\vert <e^{i \Delta \theta_{ij}}>_{\mathcal{C}_{ij}} \bigg\vert
\end{equation}
Given the great number of possible trajectories, the central limit theorem sets the distribution of the trajectories length to a Gaussian distribution. The phase $\theta_{ij}$, associated to the trajectory $ij$, follows then also a Gaussian distribution. Thus:
\begin{equation}
I_c \propto \bigg\vert e^{- 1/2 \,<(\Delta \theta_{ij})^2>_{\mathcal{C}_{ij}}} \bigg\vert
\label{Icgauss}
\end{equation}
where $\Delta\theta_{ij} = \theta_{ij} - \theta_a$ is proportional to the flux $\Phi_{ij}$ in the surface $S_{ij}$ defined by the trajectory $ij$ (Fig. \ref{semNYgilles}(d)):
\begin{equation}
	\Delta\theta_{ij} = \frac{2 e}{\hbar} \bigg[\int_i^j A_x dx - \int_1^2 A_x dx \bigg] = \frac{2 e}{\hbar} \oint A_x dx = 2 \pi \frac{\Phi_{ij}}{\Phi_0}
\label{DeltaTheta}
\end{equation}
Introducing eq.\ref{DeltaTheta} in eq.\ref{Icgauss}, we find that the critical current decays as a Gaussian function:
\begin{equation}
	I_c \propto  \bigg\vert e^{-\,(\Phi/\Phi_0)^2/(2\,\sigma^2) } \bigg\vert 
\end{equation}
where $\sigma^2 = [S^2/<S_{ij}^2>_{\mathcal{C}_{ij}}]/(4\,\pi^2)$, and $S=w \times L$. If the trajectories are ballistic, a Gaussian decay with $\sigma$=0.28 is found. \cite{sigma} The Gaussian fit of sample WAu-N gives $\sigma$=0.5, while in our previous results, \cite{Lionel} we obtained $\sigma$ in the range 0.74-1.8 for aspect ratios L/w in the range 1.9-10.4. \\
The exact calculation describing diffusive trajectories from point $i$ to point $j$ gives: \cite{Montambaux}
\begin{equation}
	I_c = I_c(0) \, \frac{\frac{\pi}{\sqrt{3}} \frac{\Phi}{\Phi_0} }{\sinh \bigg(\frac{ \pi}{\sqrt{3}} \frac{\Phi}{\Phi_0}\bigg)}
\label{gilles}
\end{equation}
The difference between the above distribution and a Gaussian one is shown in the inset of Fig. \ref{IcHN}. While this model explains the origin of the quasi-gaussian shape of our $I_c(\Phi)$ curves, the experimental decay is faster than the predicted one. A more precise calculation, taking into account the 2D nature of both samples, is thus necessary.\\

\emph{2D Usadel equation --} J.C. Cuevas and F.S. Bergeret \cite{CuevasIcH} have solved the 2D Usadel equation for long diffusive SNS junctions with low-resistance interfaces and for different aspect ratios of the normal wire. 
A complete field penetration in the normal metal ($w<\lambda_{L,N}$) is assumed, neglecting any Josephson current screening effects. Moreover no inelastic scattering is considered.
The authors found two very different limits:\\
- $w<<L$ : in this 1D limit, the field acts as a pair-breaking mechanism. The critical current is then monotonically reduced: at T=0 $I_c(B)/I_c(0) \sim e^{-0.238\, (\Phi / \Phi_0)^2}$ (inset of Fig. \ref{IcHN}). \\
- $w>>L$ : in this case, the main effect of the field is to modulate the phase over the junction width, giving rise to an interference pattern, which is identical to a Fraunhofer pattern for small enough aspect ratios ($L/w<$0.04). \\
The authors do not suppose a purely sinusoidal current-phase relation to start with, but calculate the complete $I(\delta)$ relation of a long SNS junction. However, the wide junction limit corresponds exactly to the Fraunhofer pattern obtained in the case of a sinusoidal current-phase relation. \\

\emph{Experimental results --} In agreement with the prediction of J.C. Cuevas et al. for the aspect ratio $L/w$=4.5,
junction WAu-N does not exhibit any oscillating pattern (Fig. \ref{IcHN}). The field decay cannot however be fitted by the numerical simulation for the corresponding flux scale $\Phi/\Phi_0$: the experimental decay is faster than the predicted one by a factor 2.5. This can be explained by taking into account a non-perfect interface, \cite{CuevasInterfaces} with an interface resistance $R_i$ given roughly by $R_{i}=2 R_N \sim 5 \,\Omega$ (the ratio $R_i/R_N =2\,$ leads indeed to a rescaling of the perfect interface curve by a factor 2.3). This non-perfect interface may be attributed to the right W contact, which is just at the extremity of the gold wire (see the SEM image \ref{semNYgilles}.b).\\
To understand the experimental $I_c(\Phi)$ of sample WAu-Sq, it is necessary to take into account not only the aspect ratio of the normal wire, but also the flux correction due to the inductance of the W contacts, evident in the tilt of the central peak of the $I_c(\Phi)$ curve. The 'internal' flux $\Phi_{int}$ through the normal wire is thus the sum of the applied external flux $\Phi$ and the flux created by the current flowing in the superconducting contacts, $\Phi_{\mathcal{L}} = \mathcal{L} \times I_c(\Phi_{int})$, following the below relation.
\begin{equation}
	\Phi_{int} = \Phi + \mathcal{L}\,I_c(\Phi_{int})
\end{equation}
The experimental curve should then be compared not to the calculated curve $I_c^{th} (\Phi_{int})$ but to $I_c^{th} (\Phi) = I_c^{th} (\Phi_{int}-\mathcal{L}\,I_c^{th}(\Phi_{int}))$. 
The best fit of our data for sample WAu-Sq according to the above expression yields $\mathcal{L}$=11.5 pH. This value is compatible with the kinetic inductance, much larger than the geometrical one, of similarly made W wires.\\
When comparing our experimental results with the numerical simulation of 2D Usadel equation for $L/w=0.7$ modified by the self-inductance, we find a good agreement in the position of the zeros for $\Phi=B\,S$, with $S=3.25 \mu m^2$. This corresponds roughly to the whole surface of the normal metal square, slightly larger than the surface between the contacts. The amplitude of the measured oscillations following the central peak decreases slower than predicted, and the minima of the first periods do not go to zero; this can be qualitatively explained as the effect of a non uniform current distribution in the normal metal. \cite{Barone}\\

\begin{figure}[hb]
\includegraphics[width=\columnwidth]{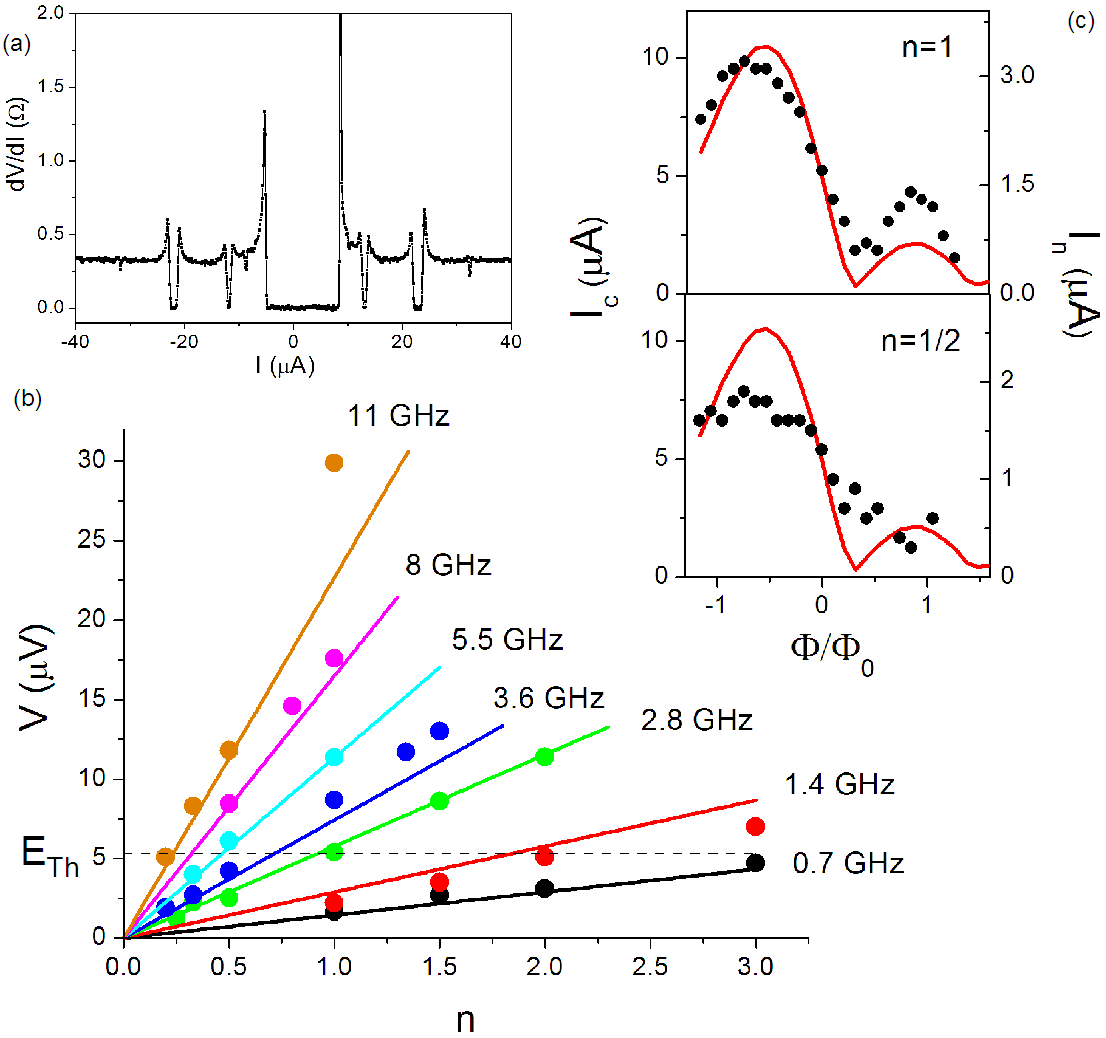}
\caption{(color online) (a) Differential resistance dV/dI of sample WAu-Sq in presence of an irradiation at frequency f=2.8 GHz. (b) Dc voltage vs. Shapiro step order for sample WAu-Sq; lines show the predicted dependence $V=h/2e \,n\, f $. (c) Normalised flux dependence of Shapiro steps current amplitude for n=1 and n=1/2 (symbols) compared to the critical current dependence under microwaves, which explain the decrease of its amplitude (continuous lines); the trapped flux of $4.3\,G$, corresponding to $0.7 \, \Phi_0$ is partially compensated by the inductance-related tilt of $I_c(B)$ central peak.}
\label{Shap}
\end{figure}

\emph{Shapiro steps -- } We have measured the differential resistance dV/dI when irradiating the junctions with microwaves from an antenna. We have observed Shapiro steps, in the form of microwave-induced zero resistance dips at $V=\frac{h}{2e} \,n \,f$. They result from the resonance of the ac voltage at frequency $f$ induced by the microwaves, with the current oscillations at frequency $f_J = 2eV/h$, due to the ac Josephson effect at finite voltage V (Fig. \ref{Shap}(a),(b)). In addition to the Shapiro steps at integer n, we find fractional Shapiro steps at n=1/2, n=1/3, n=3/2 and n=1/4. Both integer and fractional steps can be observed at voltages larger than the Thouless energy ($E_{Th}/e=5.3\, \mu $V), and for frequencies larger than the minigap ($\tilde{\Delta}/h=4 \,$GHz); the voltage was directly deduced from the measured V(I) curves. Fractional Shapiro steps appear in SNS junctions as a consequence of a non-sinusoidal current-phase relation. The additional harmonics in the current-phase relation can be generated by multiple coherent Andreev reflexions (MAR), when the coherence length is much longer than the N length, or by non-equilibrium effects. Fractional Shapiro steps reflect the behavior of each harmonic individually: the step at n=1/2, for example, is generated by the second harmonic and is proportional to its amplitude, having the same dependence on field and temperature. We have studied the magnetic field dependence of both integer and fractional steps amplitudes. As expected, the field dependence of the integer Shapiro steps follows roughly that of the zero-th order step, the critical current (Fig. \ref{Shap}(c)). This is not the case however for the fractional steps: in Fig. \ref{Shap} we show that the decay with the normalised flux of step n=1/2 is slower than that of the critical current. If this fractional Shapiro step was due to the MAR at equilibrium, the field dependence would show a periodicity half that of the critical current, corresponding to the double length covered by the Andreev pairs, and the zeros of the Fraunhofer pattern should be found at multiples of $\Phi_0/2$. This is evidently not the case, so that the supplementary harmonic in the current-phase relation may be traced back to non-equilibrium effects, as already suggested in P. Dubos {\it{et al.}} \cite{Dubos} and F. Chiodi {\it{et al.}}, \cite{Chiodi} where the effect of the magnetic field at the lowest temperatures was shown to increase the total critical current in the out of equilibrium SNS junction. \\

\emph{Conclusions --} We have measured the $I_c(\Phi)$ curves for two different geometries of long SNS junctions. The samples are made by a Au wire connected to W contacts, via FIB-assisted deposition. We have observed a monotonic gaussian-like decay for a quasi 1D normal wire, in contrast to the Fraunhofer-like interference pattern of a square normal wire. We explain the monotonic limit with a semiclassical 1D model, and fit both field dependences with numerical simulations of the 2D Usadel equation. Moreover, we have observed both integer and fractional Shapiro steps and their dependence in magnetic field. While integer steps follow as expected the field dependence of the critical current, fractional steps decay slower than $I_c$. This is incompatible with equilibrium MAR-originated steps, but may be explained as an out-of-equilibrium effect.   \\

We are grateful to M. Aprili, R. Deblock, B. Reulet, A. Chepelianskii, J. Gabelli and I. Petkovic for fruitful discussions. This work was supported by RTRA Triangle de la Physique NanoMan and ANR QuantADN.



\begin{thebibliography}{99}
\bibitem{Sandrine} The concentrations are deduced with a $50\%$ error by EDX measurement.
\bibitem{sigma} $\sigma$ in the ballistic case has been calculated from the discrete sum $<S_{ij}^2>_{\mathcal{C}_{ij}}=\frac{\Delta y}{w} \sum_{n=1}^{w/\Delta y} (n\,L\,\Delta y)^2 = \frac{L^2}{6}\, (w+\Delta y)(2w+\Delta y)\rightarrow \frac{w^2 L^2}{3}$.
\end{thebibliography}
\end{document}